\documentclass[twocolumn,showpacs,aps,floatfix,pra,superscriptaddress,nofootinbib]{revtex4}
\usepackage{color}
\usepackage{graphicx}
\usepackage{bm}
\usepackage{amsmath}
\usepackage{enumerate}
\begin{document}
\title{Time-dependent potential barriers and superarrivals}
\author{H. Karami}
\email{h.karami@stu.qom.ac}
\affiliation{Department of Physics, The University of Qom, P. O. Box 37165, Qom, Iran}

\author{S. V. Mousavi}
\email{vmousavi@qom.ac.ir}
\affiliation{Department of Physics, The
University of Qom, P. O. Box 37165, Qom, Iran}
\affiliation{ Institute for studies in Theoretical
Physics and Mathematics (IPM), P.O. Box 19395-5531, Tehran, Iran }
\begin{abstract}
Scattering of a Gaussian wavepacket from rectangular potential
barriers with increasing widths or heights is
studied numerically. It is seen that during a certain
time interval the time-evolving transmission
probability increases compared to the corresponding
unperturbed cases. In the literature this effect is known as
superarrival in transmission probability.
We present a trajectory-based explanation for this
effect by using the concept of quantum potential energy and
computing a selection of Bohmian trajectories.
Relevant parameters in superarrivals are
determined for the case that the barrier width increases linearly
during the dispersion of the wavepacket.
Nonlinear in time perturbation is also considered.
\end{abstract}
 \maketitle

\section{Introduction}
Success in numerical solution of time-dependent differential
equations in recent decades has provided a powerful tool for
studying quantum systems with time-dependent boundary. Interesting
phenomena like diffraction in time \cite{Mo-PR-1952} and
superarrivals \cite{BaMaHo-PRA-2002, MaHo-Pranama-2002,
MaMaHoPLA-2002, HoMaMa-JPA-2012} are seen in such systems with
time-dependent Hamiltonians.
Superarrivals is observed when a rectangular potential barrier is
perturbed by changing its height during the scattering of a Gaussian
wavepacket in a very short time: compared to the
unperturbed situation an enhancement in the time-dependent
transmission (reflection) probability is seen for a specific
time-interval if the barrier height is raised (reduced)
\cite{BaMaHo-PRA-2002, MaHo-Pranama-2002, MaMaHoPLA-2002}.
This phenomenon has been explained by taking the Schr\"{o}dinger
wavefunction as a real physical field: disturbance due to the
perturbed barrier propagates through this field to the measuring
apparatus. Propagation speed depends on the rate of change in
barrier height \cite{BaMaHo-PRA-2002, MaHo-Pranama-2002,
MaMaHoPLA-2002}. The origin of superarrivals has been explained by
the concept of quantum potential energy \cite{MaMaHoPLA-2002}
in Bohmian mechanics. Recently superarrivals were
studied for a parabolic potential barrier in position with a
time-dependent intensity and it was shown that this effect can
be interpreted semiclassically \cite{HoMaMa-JPA-2012}. It was argued
that this phenomenon can be used in secure transmission of
information \cite{MaHo-Pranama-2002, HoMaMa-JPA-2012}.

We aim to consider superarrivals in some more general
situations. We will proceed as follows: The occurrence of
superarrivals is shown in section \ref{sec: gauss_lw} for the
scattering of a Gaussian wavepacket by a rectangular barrier whose
width increases linearly in time.
Then we study the effect of perturbation on the
time-evolving expectation values of Hamiltonian, momentum and
position operators for the transmitted part of the wavepacket.
After, superarrivals are studied in the context of Bohm's causal theory by
computing a selection of Bohmian trajectories
and noting the concept of quantum potential energy.
Section \ref{sec: sup_general} generalizes the
problem to the case where the height of the potential barrier
increases nonlinearly from zero to a finite height. Finally, a
summary of our conclusions will be presented in section \ref{sec:
summary}.

%

\section{Gaussian wavepacket and linear increase in barrier width} \label{sec: gauss_lw}

Consider an ensemble of single-particle scattering experiments. In each trial, a particle
described by a Gaussian wavepacket $\psi_0(x)$
\begin{eqnarray} \label{eq: sai0}
\psi_0 (x) = \frac{1}{ (2 \pi \sigma_0^2)^{1/4} }~
e^{- (x -x_0)^{2}/4 \sigma_0^2 +i p_0(x-x_0)/\hbar}~,
\end{eqnarray}
is incident at $t=0$ from the left on a potential
barrier of height $V_0$. At a point $x_d$ in the right of the
barrier is an ideal detector that triggers when the particle reaches
the plane $x=x_d$. The initial centroid $x_0$ and root mean square width
$\sigma_0$ of $\psi_0(x)$ is chosen in a way that it has a
negligible overlap with the potential barrier.
Then, the time-varying transmission probability is given by
\begin{eqnarray} \label{eq: trans_probability}
T(t) &=& \int _{x_{d} }^{\infty} \vert \psi\left( x,t \right) \vert
^{2} dx ~.
\end{eqnarray}
The above study is done for both the case of
a static barrier, $V(x) = V_{0}~\theta(x)~\theta(w_i-x)$,
and also when the barrier is perturbed by increasing
its width from the initial width $w_i$ to a final
width $w_f$ linearly in time, $V(x,t) = V_{0}~\theta(x)~\theta(w(t)-x)$.
Here, $ \theta(x) $ is the step function and $ w(t) $ is the
time-dependent width of the perturbed barrier,
\begin{eqnarray*}\label{eq: width-t}
w(t) &=&
\begin{cases}
w_{i}
& \text{if } t \leq t_{p}
\\
w_{i} + \dfrac{w_{f} -w_{i}}{\varepsilon} (t - t_{p} )
& \text{if}~t_{p} < t \leq t_{p} +\varepsilon
\\
w_{f}
& \text{if } t > t_{p} + \varepsilon
\end{cases}
~,
\end{eqnarray*}
where $ t_{p} $ is the time at which perturbation is started and $
\varepsilon $ is the duration of perturbation.

\begin{figure}
\includegraphics[width=7cm,angle=-90]{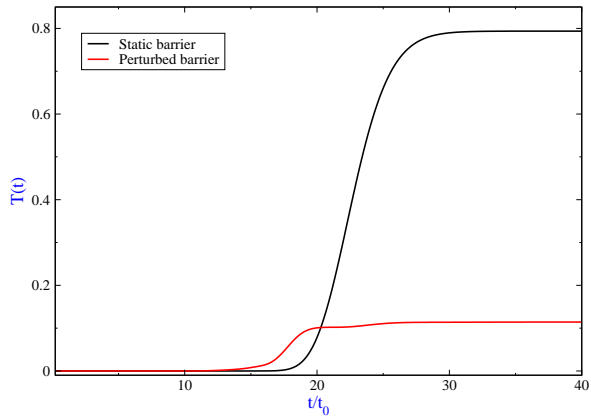}
\caption{(Color online) Time-evolving transmission probability
during the scattering of a Gaussian packet by a barrier with a
constant (black curve) and a width in linear increase (red curve).}
\label{fig: T(t)}
\end{figure}

We compute $ \psi(x,t) $ in whole space at any instant by using the
Crank-Nikolson method for numerical solution of the time-dependent
Schr\"{o}dinger equation. In this regard $ [0,40
t_{0}] $ ($t_{0} = m\sigma_{0}/p_{0}$) is taken as
time range and $ [-500 \sigma_{0} , 500 \sigma_{0}]
$ as space range. For numerical calculations we work in a system of
units where $ \hbar=1 $ and $ m=1/2 $ and parameters
are chosen in a way that spreading of the packet is negligible
during the scattering process. The constants are chosen as follows:
$ \sigma _{0} = \sigma_1/\sqrt{2}$, $ x_{0} = -6 \sigma _{1} $, $
p_{0} = 50 \pi $, $ V_{0} = 1.5 E_{0} $, $ w_{i} =0.08 \sigma_{1}$ ,
$ w_{f} =0.48 \sigma_{1}$ and $ x_{d}=10 \sigma_{1} $ where
$\sigma_1 = 0.05$. Here, $ E_{0}= p_{0}^{2}/2m +
\hbar^{2}/8m\sigma_{0}^{2} $ is the expectation value of energy for
the initial packet. Fig.~\ref{fig: T(t)} shows
time-varying transmission probability $T(t)$ for both static and
perturbed barriers. Here, perturbation takes place in time interval
$ [7.14 t_{0},7.41 t_{0}] $, that is $\varepsilon =
0.27 t_{0}$. Noting this figure one finds a finite time interval
$ \Delta t $ during which the probability of
transmission in the perturbed case is greater than the corresponding
value for the unperturbed case (superarrivals in transmission
probability). This means in the perturbed case it is
possible to find the particle beyond the detector after a shorter
time, although the overall probability of finding the particle
beyond the detector is much suppressed; $\lim_{t \rightarrow \infty}
T_s(t) \simeq 0.79$ while $\lim_{t \rightarrow \infty} T_p(t) \simeq
0.11$. Here and in the following the subscript "$s$" ("$p$")
stands for the static (perturbed) situation. Following
\cite{BaMaHo-PRA-2002} we show the time interval of
early arrivals by $ \Delta t = t_{c} - t_{d} $ where $ t_{c} $ is
the instant when the two curves cross and $ t_{d} $ is the time when
their deviation starts. From figure \ref{fig: T(t)}
one finds $t_d \simeq 10.41 t_0 $ and $t_c \simeq 20.29 t_0 $.

\begin{figure}
\centering
\includegraphics[width=7cm,angle=-90]{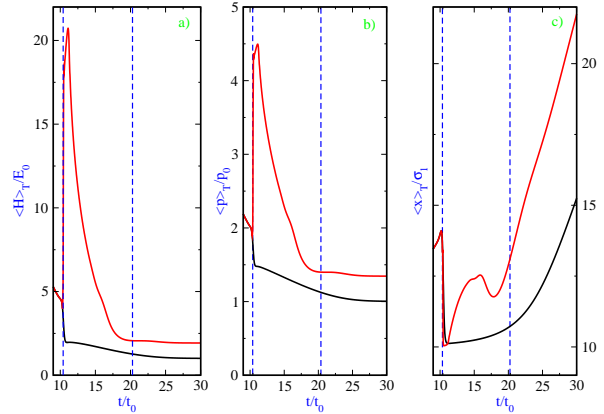}
\caption{ (Color online) Expectation value of (a) Hamiltonian, (b)
momentum and (c) position with respect to the transmitting
wavepacket. In all plots black curve shows the static situation
while the red one stands for the perturbed barrier.
Vertical blue dashed lines show $t_d$ and $t_c$.}
\label{fig: observables}
\end{figure}

To see the origin of these early arrivals we
examine mean value of observables. Expectation value of an
observable $\hat{A}$ with respect to the {\it transmitting} part of the
packet is given by
\begin{eqnarray*}
\langle \hat{A} \rangle_{\text{T}} (t) &=& \frac{ \int _{x_{d} }^{\infty} \psi^{*}( x,t ) ~ A(x) ~ \psi(x,t) dx } { T(t) }
~, \label{eq: hbar-t}
\end{eqnarray*}
where subscript "T" stands for transmission and $A(x)$ is the
observable $\hat{A}$ in the position representation.
Due to a kick imparted by the perturbed barrier,
transmitted packet moves faster in the perturbed situation in
comparison to the unperturbed case. As a result the mean energy and
momentum of transmitted packet for the perturbed barrier exceeds
those for the static case. See Fig.~\ref{fig: observables}. This leads
to the sooner arrival of the particles at the detector place.
Deviation of perturbed and static curves takes
place at $ t_d $ in agreement with that of Fig.~\ref{fig:
T(t)}. Asymptotic values of $\langle \hat{H} \rangle_{T}$ and
$\langle \hat{p} \rangle_{T}$ are respectively  $1.92 E_0$ and $1.34
p_0$ for the perturbed situation. In this limit $\langle \hat{x}
\rangle_{T}$ moves with a constant velocity.

As a measure of early arrivals the quantity
\begin{eqnarray} \label{eq: eta}
\eta &=& \dfrac{I_{p}-I_{s}}{I_{s}} ~,
\end{eqnarray}
has been defined \cite{BaMaHo-PRA-2002}, where $ I_{p} $ and $ I_{s}
$ are respectively the area under the curves of $ T_{p} (t) $ and $
T_{s} (t) $ during the time interval $\Delta t$:
\begin{eqnarray} \label{I_{p}}
I_{p} &=& \int _{\Delta t} T_{p} (t) dt~,~~~~~~~~I_{s} = \int _{\Delta t} T_{s} (t) dt~.
\end{eqnarray}
The magnitude $\eta$ of superarrivals has been plotted for three different
values of barrier height $ V_{0} $ versus the duration of
perturbation $ \varepsilon $ in figure \ref{fig:
eta_width_linear(t)}a).
%
\begin{figure}
\includegraphics[width=7cm,angle=-90]{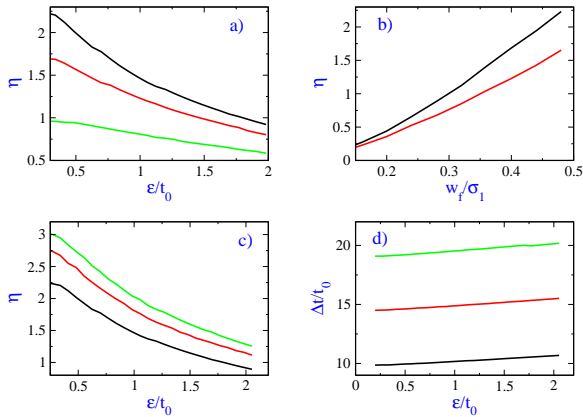}
\caption{(Color online) a) Superarrivality $\eta$ versus duration of
perturbation $\varepsilon$ for three different values of barrier
height: $V_0 = E_0$ (green curve), $V_0 = 1.3 E_0$
(red curve) and $V_0 = 1.5 E_0$ (black curve).
b) Superarrivality $\eta$ versus final width $w_f$ of the barrier
for two different values of duration of perturbation: $\varepsilon =
0.27 t_0$ (black curve) and $\varepsilon = 0.80 t_0$ (red curve).
c) Superarrivality $\eta$ versus duration of perturbation
$\varepsilon$ for three different values of detector location: $x_d
= 10 \sigma_1$ (black curve) and $x_d = 15 \sigma_1$ (red curve) and
$x_d = 20 \sigma_1$ (green curve).
d) Duration of superarrivals $\Delta t$ versus duration of
perturbation $\varepsilon$ for three different detector location:
$x_d = 10 \sigma_1$ (black curve) and $x_d = 15 \sigma_1$ (red
curve) and $x_d = 20 \sigma_1$ (green curve). }
\label{fig: eta_width_linear(t)}
\end{figure}
One sees that $ \eta $ decreases with $ \varepsilon $ for a given
value of $ V_{0} $ while for a given value of duration of perturbation,
superarrivality increases with the height of the barrier. In
Fig.~\ref{fig: eta_width_linear(t)}b) we have plotted $ \eta $
versus $ w_{f} $ for two different values of $ \varepsilon $. One
sees that the magnitude of superarrivals increases with the final
width of the perturbed barrier. In figures \ref{fig:
eta_width_linear(t)}c) and \ref{fig: eta_width_linear(t)}d) we have
plotted $ \eta $ and $ \Delta t $ versus $ \varepsilon $ for three
different values of detector position. According to these
plots we can say that for a given value of $
\varepsilon $, magnitude and duration of superarrivals increase when
the distance of the detector from the barrier becomes larger.
Increment of $ \Delta t $ with $ \varepsilon $ is gradual for a
given value of $ x_{d} $.

\begin{figure}
\centering
\includegraphics[width=7cm,angle=-90]{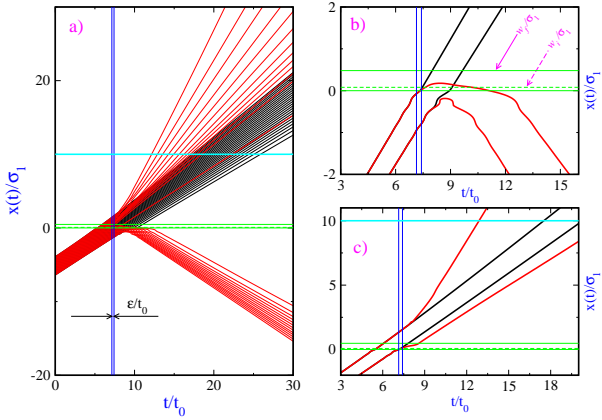}
\caption{(Color online) a) A selection of Bohmian
trajectories to have an overview of the problem, b) two typical
reflected trajectories (in perturbed situation) and c) two typical
transmitted paths. Black (red) trajectories are for the static
(perturbed) situation. Vertical blue lines show the beginning and
the end of perturbation, horizontal green lines show the borders of
the barrier and the horizontal cyan line shows the detector place.
}
\label{fig: BM_traj}
\end{figure}

Our aim is now description of early arrivals within the framework of
Bohmian mechanics (BM). In BM complete description of a system
is given by its wavefunction and position. The wavefunction which is
the solution of the Schr\"{o}dinger equation guides the particle
motion by the guidance equation,
\begin{eqnarray} \label{eq: xt}
\dot{x}(t) &=& \frac{1}{m}  \nabla S(x, t)
\bigg\vert_{x=x(t) }~,
\end{eqnarray}
where, $S$ is the phase of the wavefunction in its
polar form $\psi=R e^{iS/\hbar}$ and $x(t)$
is the particle trajectory \cite{Ho-Book-1993}. BM reproduces the results of the
standard quantum mechanics provided that distribution of initial
positions is given by $\rho = |\psi|^2$. Particle trajectories are
obtained by integrating the guidance equation (\ref{eq: xt}) for a
given initial position $x^{(0)}$. In the
second-order point of view of BM acceleration of Bohmian particle
along its trajectory is given by
\begin{eqnarray*} \label{eq: qv}
\ddot{x} &=& - \frac{1}{m} \nabla( V + Q) \bigg\vert_{x=x(t)}~,
\end{eqnarray*}
where the particle is subjected to a quantum force $ -\nabla Q $ in addition to the classical force $ -
\nabla V $ \cite{Ho-Book-1993}.
$Q$ is called quantum potential and is given by
\begin{eqnarray*} \label{eq: q}
Q (x,t) &=& - \dfrac{\hbar^{2}}{2m} \frac{\nabla^2 R}{R}~,
\end{eqnarray*}
where R is the amplitude of the wavefunction.
Due to the non-crossing property of Bohmian trajectories,
there is a critical trajectory (starting at $x_c^{(0)}$) that separates
transmitted trajectories from the reflected ones in a scattering process
and is given by \cite{Le-PLA-1993}
\begin{eqnarray} \label{eq: cr_traj}
\lim_{t \rightarrow \infty}T(t) &=& \int_{x_c^{(0)}}^{\infty} dx ~
|\psi_0(x)|^2 = \frac{1}{2} {\text{Erfc}} \left(
\frac{x_c^{(0)}-x_0}{\sqrt{2} \sigma_0} \right)~.
\end{eqnarray}
From our results for the asymptotic value of transmission
probability, we obtain from Eq. (\ref{eq: cr_traj}) $ x_c^{(0)}
\simeq x_0 - 0.82 \sigma_0 = -6.58 \sigma_1$ for the static barrier
and $ x_c^{(0)} \simeq x_0 + 1.2 \sigma_0 = -5.15 \sigma_1$ for the
perturbed one.
We have plotted in Fig.~\ref{fig: BM_traj}a) a selection of Bohmian
paths with a starting point in the range $x^{(0)} > -6.58
\sigma_1$.
Whit this condition all paths are eventually transmitted in the
static case while in the perturbed barrier in place there are two
groups of trajectories: (i) reflected ones with an initial position
in the range $ -6.58 \sigma_1 < x^{(0)} < -5.15 \sigma_1 $ and
(ii) transmitted trajectories with an initial position in the range $
-5.15 \sigma_1 < x^{(0)} $. Each of these two groups splits in two
sub-groups: (a) some reflected trajectories never reach the barrier
while a few ones reach and penetrate, but eventually turn around
(b) most transmitted particles are accelerated with respect to the
static case and produce earlier arrivals while a few ones are
decelerated and thus arrive in detector later than the corresponding
paths for the static case. See Figs.~\ref{fig: BM_traj}b) and \ref{fig:
BM_traj}c) for typical such paths. In summary, the
effect of the perturbation is to reflect more trajectories and to
push those that manage to pass the barrier. In
this connection the quantum potential $Q$ plays a crucial rule in
propagating the influence of barrier perturbation far from where the
barrier is non-zero.


\section{Gaussian wavepacket and nonlinear increase in barrier height} \label{sec: sup_general}

\begin{figure}
\centering
\includegraphics[width=7cm,angle=-90]{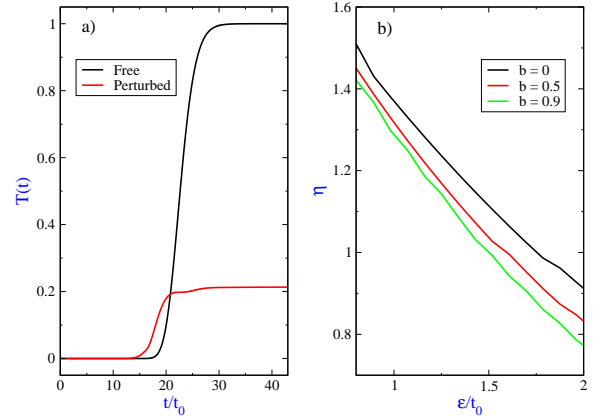}
\caption{(Color online) Scattering of a Gaussian packet by a barrier
with a height in nonlinear increase: a) Time
evolution of transmission probability for the free (black curve) and
perturbed case (red curve) with $a=0.1$ and $b=0.9$. b)
Superarrivality versus duration of perturbation for three different
values of non-linear coefficient.}
\label{fig: height_nonlinear(t)}
\end{figure}

It has been shown that during a finite time interval
the time-varying probability of transmission
exceeds that for free propagation in the scattering of a Gaussian
wavepacket by a barrier with a height in linear increase
\cite{MaMaHoPLA-2002, MaHo-Pranama-2002}. As the first
generalization, we consider a situation where the height of the
rectangular barrier changes nonlinearly in time from
$0$ to $V_0$ as follows,
\begin{eqnarray} \label{eq: vunlinear}
V (x,t) &=& V_{0} \hspace{1mm}\theta (x+\dfrac{w}{2}) \hspace{1mm}\theta (\dfrac{w}{2} -x)
\nonumber \\
& \times &
\begin{cases}
0 & \text{if} \hspace{1mm} t\leq t_{p} \\
a\left( \dfrac{t-t_{p}}{\varepsilon} \right) +b\left( \dfrac{t-t_{p}}{\varepsilon} \right) ^{2} & \text{if}
\hspace{1mm} t_{p}< t \leq t_{p} +\varepsilon \\
1 & \text{if} \hspace{1mm} t > t_{p} + \varepsilon
\end{cases}
\nonumber \\
\end{eqnarray}
From Eq. (\ref{eq: trans_probability}) one obtains
\begin{eqnarray} \label{eq: free_trans}
T_{\text{f}} &=& \frac{1}{2} \left \{1+ {\text{Erf}}{\left( \frac{p_0~t/m+x_0-x_d}{\sqrt{2}\sigma_t} \right)} \right \}~,
\end{eqnarray}
for the time-varying transmission probability in free propagation,
where $\sigma_t = \sqrt{1+ ( \hbar t/2m\sigma_0^2 )^2}$ is the rms
width of the time-evolving wavepacket. In our calculations we have
imposed constraints $ a+b=1 $, $ 0\leq a\leq1 $ and $ 0\leq b\leq1
$. In Fig.~\ref{fig: height_nonlinear(t)}a) we have plotted time-evolving
transmission probability for
$a=0.1$ and $b=0.9$. Perturbation takes place during the time
interval $ [7.14 t_0, 7.41 t_0] $ where the height of
barrier increases nonlinearly from $ 0 $ to $ 2 E_{0} $ and we have
put $x_d=10\sigma_1$ and $ w=0.32 \sigma_{1} $.

In order to show dependence of superarrivals on the
nonlinear coefficient $b$, we have plotted $ \eta $ versus $
\varepsilon $ in figure \ref{fig: height_nonlinear(t)}b)
for three different values of $b$. As shown in
this figure $ \eta $ decreases with $ \varepsilon $ ($b$) for a
given value of $b$ ($ \varepsilon $). When the height
of the barrier increases in the nonlinear form (\ref{eq:
vunlinear}), the rate of the change of barrier's
height is
$V_0[\frac{a}{\varepsilon}+\frac{2(1-a)}{\varepsilon^2}(t-t_p)]$
 in contrast to the constant rate $\frac{V_0}{\varepsilon}$ in the
case of linear increase.
Thus for $ t < t_p+\frac{\varepsilon}{2} $ ($ t > t_p + \frac{\varepsilon}{2} $)
the rate of increase is higher (lower) in the case of the non-linear
perturbation than for the linear one. 
This means that in the first
half of the perturbation when the incident packet has considerable
interaction with the barrier, the potential changes
slower compared to the linear increase and thus the
kick the packet receives is weaker. As a result
superarrivals are suppressed compared to the linear increase.

At the end we just briefly provide our numerical
results in two more general cases: (i) in the scattering of a
Gaussian wavepacket from two successive rectangular potential
barriers which are perturbed by simultaneous increase in height,
magnitude of superarrivals decreases with separation of barriers and
duration of their perturbation, while increases with the final
height of the barriers (ii) in the scattering of the non-Gaussian
wavepacket \cite{ChHoMaMoMoSi-CQG-2012}
\begin{eqnarray} \label{eq: non-gau}
\psi_{0}(x) &=& \frac{ 1+\alpha\sin\left(\frac{\pi (x-x_{0})}{
4\sigma_{0}} \right) } { \sqrt{\sqrt{2\pi\sigma_{0}^{2}} \left[
1+\alpha^{2} e^{-\pi^{2}/16} \sinh(\dfrac{\pi^{2}}{16}) \right] } }
\nonumber \\
&\times & \exp \left[
{-\dfrac{(x-x_{0})^{2}}{4\sigma_{0}^{2}}+i\dfrac{p_{0}}{\hbar}(x-x_{0})}
\right]~,
\end{eqnarray}
from a barrier with a height in linear increase, magnitude of
superarrivals decreases with the duration of perturbation but does
not have regular behavior with the non-Gaussian coefficient $\alpha$.

%
\section{summary and discussion} \label{sec: summary}
In this paper we studied superarrivals in the scattering
of a wavepacket from time-dependent
rectangular potential barriers.
We showed that superarrivals in transmission probability
occurs in the scattering of a Gaussian wavepacket from a
rectangular potential barrier with a width in linear increase during a
finite time interval. Moreover, we depicted that
the magnitude of superarrivals decreases with the duration of perturbation
while grows when the final width of
the perturbed barrier or detector distance from the barrier increases.
By calculating the time evolution of Hamiltonian, momentum and
position expectation values, we depicted that when the barrier's
width increases, the velocity of transmitted packet increases
and yields superarrivals in transmission probability.
We saw the effect of the perturbation is to reflect
more trajectories and to push those that manage to pass the barrier
by computing a selection of Bohmian trajectories.

We saw when the height of the barrier increases nonlinearly, the
magnitude of superarrivals decreases with the nonlinear coefficient.
From the above studies one sees that irrespective of the shape of
the incident wavepacket (Gaussian/non-Gaussian) and the
form of perturbation (linear/nonlinear),
superarrivality decreases with the duration of the perturbation. The
reason is as follows. Larger values of $\varepsilon$
correspond to slower changes in barrier. Thus, incident wavepacket
will see a small change in the height/width of the barrier during its
interaction with the barrier. As a result transmission probability
will not be very different from that of the static barrier and thus
superarrivality diminishes. This situation
characterizes adiabatic limit in quantum mechanics.
%
%
\section*{Acknowledged}
The authors would like to acknowledge reviewers for
valuable comments on an earlier draft of the paper. We thank M. R.
Mozaffari for help in numerical calculations and S. M. Fazeli for
helpful discussions. Financial support from the University of Qom is
acknowledged.
%

\end{document}